\documentclass[11pt]{article}
\usepackage{euscript}
\usepackage{amssymb}
\usepackage{amsfonts}
\usepackage{amsbsy}
\usepackage{epsfig}
\usepackage{amsthm}
\usepackage{amscd}
\usepackage{amstext}
\usepackage{verbatim}
\usepackage{amsmath}
\usepackage{hyperref}
\usepackage{cite}

\begin{document}

\begin{center}
{\Large \textbf{Holographic renormalization \\ by Hamilton-Jacobi formulation with generated ansatz}}

\vspace{1cm}

Ming-Xia Ma$^{1}$ and Shao-Feng Wu$^{1,2}$

\vspace{1cm}

$^{1}${\small \textit{Department of physics, Shanghai University, Shanghai,
200444, China }}\\[0pt]

$^{2}${\small \textit{Center for Gravitation and Cosmology, Yangzhou
University, Yangzhou 225009, China }}\\[0pt]
\vspace{0.5cm}

{\small \textit{E-mail: mingxiama@shu.edu.cn, sfwu@shu.edu.cn}}
\end{center}

\vspace{1cm}

\begin{abstract}
In AdS/CFT corresponding, the UV divergence of generating functional on the
field theory can be removed as the IR divergence in the gravity. This
geometric process is well known as holographic renormalization. The standard
method of holographic renormalization is based on the Fefferman-Graham
expansion, which is strict and universal but technically cumbersome. To
improve the technique, different methods have been proposed. Here we
develop an alternative approach to holographic renormalization based on the
Hamilton-Jacobi formulation of gravity. Compared to previous approaches, its
distinguishing feature is the generation of exact ansatz of counterterms. We
apply this approach to several typical holographic models, which
consistently performs well.
\end{abstract}
\pagebreak

\section{Introduction}

Anti-de Sitter/conformal field theory (AdS/CFT) corresponding not only
provides a gravitational lens for the strongly coupled quantum many-body
system but also inspires the theory of quantum gravity \cite{HongLiu2004}.
Holographic renormalization (HR), which removes the UV divergence on the
boundary field theory by isolating the IR divergence in the bulk gravity, is
one of the essential components in the AdS/CFT \cite%
{Skenderis0209,Papadimitriou2016}.

There are various ways to perform the HR \cite%
{Henningson9806,Balasubramanian9902,Kraus9906,deHaro0002,Bianchi0112,deBoer9912,deBoer0101,Kalkkinen0103,Martelli0205,Papadimitriou0404,Papadimitriou0407,Papadimitriou1106,Olea0504,Olea0610,Bzowski1612,Elvang1603,Wu1903}%
. Among others, the standard method is strict, universal and conceptually simple \cite%
{Henningson9806,deHaro0002,Bianchi0112}. However, its core component, the
Fefferman-Graham (FG) expansion (especially its inversion), is technically
cumbersome \cite{Fefferman1985}, which can be partially attributable to the
breaking of covariance in the procedure. The problem of covariance is absent
in a class of approaches based on the Hamilton-Jacobi (HJ) formulation of
gravity, where the radial coordinate plays the role as time. This class of
approach was first proposed by de Boer, Verlinde, and Verlinde (dBVV) \cite%
{deBoer9912,deBoer0101}, who derive the counterterms in the derivative
expansion by iteratively solving the radial HJ equation. Note that here the HJ
equation is reduced to the Hamiltonian constraint which ensures the
invariance under the radial diffeomorphism. The dBVV's approach is
considerably improved by Kalkkinen, Martelli and Muck \cite%
{Kalkkinen0103,Martelli0205}. In particular, the logarithmic counterterms
that have not been explicitly obtained in \cite{deBoer9912,deBoer0101} are
isolated by relating them to the breakdown of the recursive equation. The
main flaw common to these works is the requirement to postulate an ansatz
consisting of all potential divergent terms. Since usually the ansatz is
constrained only to be local and covariant, it is likely to contain plenty
of redundancy while sometimes the sufficient ansatz is difficult to be
figured out. In Ref. \cite{Papadimitriou0404,Papadimitriou0407},
Papadimitriou and Skenderis put forward a systematic method without relying
on the ansatz. The crucial difference is that the covariant expansion is
organized according to the eigenfunctions of the dilatation operator
comprised of induced metric and scalar fields. However, as pointed out in
\cite{Papadimitriou1106}, the eigenfunctions of the dilatation operator for
an arbitrary scalar potential would not serve as a practical basis for the
expansion since they would be highly nontrivial\footnote{%
The dilatation operator can be understood as the asymptotic form of the
functional representation of the radial derivative. In Ref. \cite%
{Papadimitriou0404}, it has been noted that the functional representation
must be modified when the leading asymptotics of bulk fields are of the form
$r\exp (-dr/2)$. This happens when the scaling dimension of dual operators
is $\Delta =d/2$. The well-known FGPW model is exactly this case \cite%
{Freedman9904}.}. This problem has been addressed in \cite{Papadimitriou1106}%
, where the dilatation operator is replaced by the operator relevant only to
the induced metric and usually the resulting recursive equations are the
functional differential equations.

In 2016, Evang and Hadjiantonis proposed a practical approach to the HJ
formulation of HR, which returns to the derivative expansion and the
postulation of ansatz \cite{Elvang1603}. In contrast to previous approaches
for solving the Hamiltonian constraint, this one preserves the radial
partial-derivative term of the HJ equation\footnote{%
This idea is partially inspired by Ref. \cite{Larsen0307}, where the HJ
equation is used to isolate the infrared divergence of the scalar field in a
fixed de-Sitter background.}. We hence refer the former as Hamiltonian
approaches and the latter HJ approach. The HJ approach has been illustrated
in several Einstein-scalar theories, including the FGPW model, the
dilaton-axion system with constant potential and others. Interestingly, it
handles power and logarithmic divergences in completely the same way, that is, there is no difference between their derivations such as the breakdown of recursive
equations. In Ref. \cite{Wu1903}, it is clarified that the HJ approach is
not conflicted with the Hamiltonian constraint, because there only a part of
the HJ equation is actually used to perform HR while the Hamiltonian
constraint is not used. Thus, the HJ approach is strict and applicable to
the theories with or without the diffeomorphism symmetry.

In this paper, we will develop a new version of the HJ approach to HR. The
key improvement we will make is a systematic method to generate the exact
ansatz. By \textquotedblleft exact\textquotedblright , it means no omission
and no redundancy. Moreover, it will be illustrated that a general solution
to the coefficients in the ansatz can be derived. In Appendix, we will
provide some technical details and apply the new approach to some typical
holographic models. In particular, we will study the non-linear holographic
axion model, which has interesting applications in the duality between
gravity and quantum matter \cite{Taylor1406,Baggioli2014,Matteo2021,Li2108}. Since
its action allows for the derivative term with non-integral powers, this model might be a
challenge for previous HR approaches that require a complete list of all
possible ansatz or similar information. In fact, this was the original
motivation for developing our new approach.

\section{Benchmark}

We will develop the benchmark of the approach in terms of the\ Einstein
gravity coupled to scalars fields in the d+1-dimensional asymptotic AdS
spacetime. The bulk action of the theory\ is%
\begin{equation}
S=-\int_{M}d^{d+1}x\sqrt{g}\left( \mathcal{R}-g^{\mu \nu }G_{IJ}\partial
_{\mu }\Phi ^{I}\partial _{\nu }\Phi ^{J}-V\right) ,  \label{action}
\end{equation}%
where $V$ is a potential of scalar fields and $G_{IJ}$ denote their
symmetric couplings.

\subsection{HJ formulation}

Let's introduce the HJ equation of this gravity system. Using Eq. (\ref%
{action}) and the Arnowitt-Deser-Misner (ADM) metric with FG gauge \cite{ADM}%
\begin{equation}
ds^{2}=dr^{2}+\gamma _{ij}dx^{i}dx^{j},
\end{equation}%
one can obtain the radial Hamiltonian \cite{Elvang1603,Wu1903}
\begin{equation}
H=\int_{\partial M}d^{d}x\left[ \frac{1}{\sqrt{\gamma }}\left( \pi _{ij}\pi
^{ij}-\frac{1}{d-1}\pi ^{2}+\frac{1}{4}G^{IJ}\pi _{I}\pi _{J}\right) +\sqrt{%
\gamma }L_{S}\right] ,  \label{H}
\end{equation}%
where $\pi _{ij}$ and $\pi _{I}$ are the canonical momenta conjugate to the
induced metric $\gamma ^{ij}$ and the scalar fields $\Phi ^{I}$. We would
like to refer $L_{S}$ as the counterterm seed due to its role played in the
HR, which is given by
\begin{equation}
L_{S}=R-\gamma ^{ij}G_{IJ}\partial _{i}\Phi ^{I}\partial _{j}\Phi
^{J}-V(\Phi ).
\end{equation}%
In classical mechanics, the canonical momenta can be expressed as the
variation of the on-shell action $S_{\mathrm{os}}$ \cite{Landau1987}. As a
result, the HJ equation can be written as\footnote{%
For the theory with diffeomorphism symmetry, the Hamiltonian constraint $H=0$
should be respected. Thus, one might wonder why the partial derivative $%
\partial S_{\mathrm{os}}/\partial r$ is kept in general. This issue has been
explained carefully in section 2 of Ref. \cite{Wu1903}. Here we emphasize
that Eq. (\ref{HJ0}) always holds no matter the theory has the
diffeomorphism symmetry or not.
\par
\bigskip
\par
\bigskip}%
\begin{equation}
H\left( \gamma _{ij},\Phi ^{I};\frac{\delta S_{\mathrm{os}}}{\delta \gamma
_{ij}},\frac{\delta S_{\mathrm{os}}}{\delta \Phi ^{I}}\right) +\frac{%
\partial S_{\mathrm{os}}}{\partial r}=0.  \label{HJ0}
\end{equation}

In section A of Appendix, we have reviewed that the HJ equation can be
divided into two parts. We focus on the counterterm part%
\begin{equation}
H_{\mathrm{ct}}+\frac{\partial S_{\mathrm{ct}}}{\partial r}=0,  \label{CPHJ0}
\end{equation}%
where%
\begin{equation}
H_{\mathrm{ct}}=-\int_{\partial M}d^{d}x\left[ \{S_{\mathrm{ct}},S_{\mathrm{%
ct}}\}+\sqrt{\gamma }L_{S}\right] ,
\end{equation}%
\begin{eqnarray}
\{S_{\mathrm{ct}},S_{\mathrm{ct}}\} &=&\frac{1}{\sqrt{\gamma }}\left( \frac{%
\delta S_{\mathrm{ct}}}{\delta \gamma _{ij}}\frac{\delta S_{\mathrm{ct}}}{%
\delta \gamma _{kl}}\gamma _{ijkl}+\frac{1}{4}G^{IJ}\frac{\delta S_{\mathrm{%
ct}}}{\delta \Phi ^{I}}\frac{\delta S_{\mathrm{ct}}}{\delta \Phi ^{J}}%
\right) , \\
\gamma _{ijkl} &=&\gamma _{ik}\gamma _{jl}-\frac{1}{d-1}\gamma _{ij}\gamma
_{kl}.
\end{eqnarray}

For later use, we rewrite the counterterm as%
\begin{equation}
S_{\mathrm{ct}}=-2\int_{\partial M}d^{d}x\sqrt{\gamma }U(\gamma ^{ij},\Phi
^{I},r).  \label{U}
\end{equation}%
For the sake of brevity, we define%
\begin{equation}
K=4Y_{ij}Y^{ij}-\frac{1}{d-1}(U-2Y)^{2}-U^{2},  \label{EC}
\end{equation}%
\begin{equation}
Y_{ij}=\int \frac{\delta U}{\delta \gamma ^{ij}},\qquad Y^{ij}=-\int \frac{%
\delta U}{\delta \gamma _{ij}},\qquad Y=\gamma ^{ij}\int \frac{\delta U}{%
\delta \gamma ^{ij}},\qquad P_{I}=\int \frac{\delta U}{\delta \Phi ^{I}},
\label{Yij}
\end{equation}%
\begin{equation}
\int \delta U=\frac{1}{\sqrt{\gamma }}\int_{\partial M}d^{d}x\sqrt{\gamma }%
\delta U.  \label{deltab}
\end{equation}%
Then Eq. (\ref{CPHJ0}) can be reshaped as%
\begin{equation}
2\frac{\partial U}{\partial r}+L_{S}+K+G^{IJ}P_{I}P_{J}=\mathrm{0},
\label{CPHJ1}
\end{equation}%
which holds up to total derivatives, since it should be understood as an
integral equation.

\subsection{Generation of ansatz}

One of the key points in various HR approaches is how to organize the
counterterms. Considering that the counterterms are built from some boundary
invariants with different degrees of divergence, we choose to expand them
according to these degrees of divergence, which can be formally written as%
\begin{equation}
U=U_{(k_{0})}+U_{(k_{1})}+U_{(k_{2})}+\cdots .  \label{Uki}
\end{equation}%
Some remarks on the expansion are in order. First, we suppose that $%
U_{(k_{0})}$ is independent with boundary fields. This is true at least for
asymptotic AdS spacetimes. Second, the divergence
degree $k_{i}$ is defined by the asymptotic behavior of $U_{(k_{i})}$, which
can be expressed as $U_{(k_{i})}$ $\sim e^{-k_{i}r}$. Although $k_{i}$ is
often a number, it is not necessary\footnote{%
Whether $k_{i}$ is a number or not, we can use $U_{(k_{i})}$ $\sim
e^{-k_{i}r}$ to sort the counterterms and therefore Eq. (\ref{Uki}) is
always well defined. In Appendix D.2, we will deal with the FGPW model where
the divergence degree related to the scalar with $\Delta =d/2$ is not a
number.}. Third, we do not need to postulate the concrete ansatz of each $%
U_{(k_{i})}$, nor do we specify its divergence degree $k_{i}$ in advance.
Both of them will be emergent. Fourth, we assume a variation identity%
\footnote{%
This identity has been found before in different systems \cite%
{Elvang1603,Wu1903}. A similar identity has been used in \cite%
{Papadimitriou1106} and it can be traced back to \cite{Kraus9906} for pure
gravity.}%
\begin{equation}
Y_{(k_{i})}=\overline{k}_{i}U_{(k_{i})}+\mathrm{TD}\text{ with }k_{i}\geq 2%
\overline{k}_{i},  \label{VID}
\end{equation}%
where $\overline{k}_{i}$ is the number of inverse metrics in $U_{(k_{i})}$
and $\mathrm{TD}$ means total\ derivatives. In section B of Appendix, we
will prove that Eq. (\ref{VID}) holds very generally.

Now let's expand Eq. (\ref{CPHJ1}). Substituting Eq. (\ref{Uki}) into Eq. (%
\ref{CPHJ1}), we have%
\begin{equation}
2\frac{\partial }{\partial r}\sum\limits_{i}U_{(k_{i})}+\sum%
\limits_{i}L_{S(k_{i})}+\sum\limits_{m,n}H_{(k_{m},k_{n})}=0,  \label{ULH1}
\end{equation}%
where%
\begin{eqnarray}
H_{(k_{m},k_{n})}
&=&4Y_{(k_{m})ij}Y_{(k_{n})}^{ij}+G^{IJ}P_{I(k_{m})}P_{J(k_{n})}  \notag \\
&&-\frac{1}{d-1}%
(U_{(k_{m})}-2Y_{(k_{m})})(U_{(k_{n})}-2Y_{(k_{n})})-U_{(k_{m})}U_{(k_{n})}.
\end{eqnarray}

Since the variation of $U_{(k_{0})}$ vanishes, the leading order of Eq. (\ref%
{ULH1}) is reduced to:%
\begin{eqnarray}
&&2\frac{\partial U_{(k_{0})}}{\partial r}-\frac{1}{d-1}%
(U_{(k_{0})}-2Y_{(k_{0})})^{2}-U_{(k_{0})}^{2}+G^{IJ}P_{I(k_{0})}P_{J(k_{0})}+L_{S(k_{0})}
\notag \\
&=&2\frac{\partial U_{(k_{0})}}{\partial r}-\frac{d}{d-1}%
U_{(k_{0})}^{2}+d(d-1)=0,
\end{eqnarray}%
which has the solution $U_{(k_{0})}=1-d+\mathcal{O}(e^{-dr})$. Note that the
higher order $\mathcal{O}(e^{-dr})$ does not correspond to a real divergent
term since $\sqrt{\gamma }e^{-dr}\sim \mathcal{O}(1)$. With $U_{(k_{0})}$ in
hand, one can rewrite any other order of Eq. (\ref{ULH1}) as%
\begin{equation}
2\frac{\partial }{\partial r}U_{(k_{i})}+2(d-2\overline{k}%
_{i})U_{(k_{i})}+L_{S(k_{i})}+\sum\limits_{m,n\rightarrow
i}H_{(k_{m},k_{n})}=0,  \label{ULH2}
\end{equation}%
where $m,n\rightarrow i$ denotes that the choice of $m$ and $n$ in the range
$(0,i]$ such that $H_{(k_{m},k_{n})}$ has the divergence degree $k_{i}$.
Note that in derivation of Eq. (\ref{ULH2}), we have used the variation
identity (\ref{VID}).

To proceed, we set%
\begin{equation}
U_{(k_{i})}=\sum_{a}C_{(k_{i})}^{(a)}(r)\bar{U}_{(k_{i})}^{(a)},
\end{equation}%
where $\bar{U}_{(k_{i})}^{(a)}$ are some independent scalars made of
boundary fields and their derivatives. The coefficients\ $%
C_{(k_{i})}^{(a)}(r)$ are assumed to depend on $r$ in general, following
Ref. \cite{Elvang1603,Wu1903}. Without loss of generality, we will focus on
one term $C_{(k_{i})}(r)\bar{U}_{(k_{i})}$, where we have suppressed the
index $(a)$ for brevity. Then Eq. (\ref{ULH2}) indicates
\begin{equation}
2\frac{\partial }{\partial r}C_{(k_{i})}\bar{U}_{(k_{i})}+2(d-2\overline{k}%
_{i})C_{(k_{i})}\bar{U}_{(k_{i})}+L_{A(k_{i})}+L_{P(k_{i})}=0,  \label{UCLL}
\end{equation}%
where%
\begin{equation}
L_{A(k_{i})}=L_{S(k_{i})}+\sum\limits_{\substack{ m,n\rightarrow i  \\ %
m,n\neq i}}H_{(k_{m},k_{n})},  \label{LAK}
\end{equation}%
\begin{eqnarray}
L_{P(k_{i})} &=&\sum\limits_{\substack{ m,n\rightarrow i  \\ m=i\ or\ n=i}}%
G^{IJ}P_{I(k_{m})}P_{J(k_{n})}+\sum\limits_{\substack{ m,n\rightarrow i  \\ %
m=i\ and\ n=i}}G^{IJ}P_{I(k_{m})}P_{J(k_{n})}  \label{LPK} \\
&=&C_{(k_{i})}\bar{U}_{(k_{i})}b_{1(k_{i})}+C_{(k_{i})}^{2}\bar{U}%
_{(k_{i})}b_{2(k_{i})}.  \label{b1b2}
\end{eqnarray}%
In section C of Appendix, we will analyze two constants $b_{1(k_{i})}$ and $%
b_{2(k_{i})}$ in detail. One can find that either $b_{1(k_{i})}$ or $%
b_{2(k_{i})}$ must be zero. We will refer them as mass parameters since
usually they are related to the mass of scalar fields.

Observing Eqs. (\ref{UCLL})-(\ref{b1b2}), one may notice an important fact:
the $i$-th order counterterms are absent if $L_{A(k_{i})}=0$. This can be
seen by solving Eq. (\ref{UCLL}) without $L_{A(k_{i})}$:%
\begin{equation}
2\frac{\partial }{\partial r}C_{(k_{i})}+2(d-2\overline{k}%
_{i})C_{(k_{i})}+b_{1(k_{i})}C_{(k_{i})}+b_{2(k_{i})}C_{(k_{i})}^{2}=0.
\label{Cbb}
\end{equation}%
Its solutions can be divided into three classes.

1. For $b_{1(k_{i})}=b_{2(k_{i})}=0$, the solution is%
\begin{equation}
C_{(k_{i})}=\alpha e^{-(d-2\bar{k}_{i})r}+\cdots .
\end{equation}

2. For $b_{1(k_{i})}\neq 0$ and $b_{2(k_{i})}=0$, the solution is%
\begin{equation}
C_{(k_{i})}=\alpha e^{-(d-2\bar{k}_{i}+b_{1(k_{i})}/2)r}+\cdots .
\end{equation}

3. For $b_{1(k_{i})}=0$ and $b_{2(k_{i})}\neq 0$, the solution is%
\begin{equation}
C_{(k_{i})}=\left\{
\begin{array}{c}
\alpha e^{-(d-2\overline{k}_{i})r}+\cdots ,\ d-2\overline{k}_{i}\neq 0 \\
\frac{2}{b_{2(k_{i})}r}+\cdots ,\ d-2\overline{k}_{i}=0%
\end{array}%
\right. .  \label{C32}
\end{equation}

Here $\alpha $ is an arbitrary integral constant. Immediately, this implies
that the solutions with $\alpha $ are not relevant to any real divergent
terms, because the divergent part of counterterms should be unique\footnote{%
In this work, we are not concerned with finite counterterms. They depend on
the choice of RG scheme.}. Moreover, the second line of Eq. (\ref{C32}) does
not yield real divergent terms either, since%
\begin{equation}
\sqrt{\gamma }C_{(k_{i})}\bar{U}_{(k_{i})}\sim e^{dr}\cdot \frac{1}{r}\cdot
e^{-k_{i}r}=\frac{e^{-(k_{i}-2\overline{k}_{i})r}}{r}<\mathcal{O}(1),
\end{equation}%
where we have used $k_{i}\geq 2\overline{k}_{i}$.

As a result, one can see that the ansatz $U_{(k_{i})}$ at order $i$ should
be emergent in $L_{A(k_{i})}$. Put differently, the divergence degree and
the ansatz at each order can be iteratively generated. To be more clear, we
define the ansatz generator%
\begin{equation}
L_{A}=L_{S}+\sum\limits_{m,n}H_{(k_{m},k_{n})}.  \label{LAG}
\end{equation}%
One should input the counterterm seed of order greater than $i-1$ and the ansatz of
order in the range $(0,i-1]$. The $L_{A(k_{i})}$ defined by Eq. (\ref{LAK})
can be identified as the term in $L_{A}$ with $k_{i}$ closest to $k_{i-1}$
but greater than it.

\subsection{Solution of coefficients}

We will specify the coefficients in the ansatz by solving Eq. (\ref{UCLL}).
To save symbols, let's still denote any term in $L_{A(k_{i})}$ as $\bar{U}%
_{(k_{i})}$ and the ansatz as $C_{(k_{i})}\bar{U}_{(k_{i})}$. Thus, Eq. (\ref%
{UCLL}) can be reduced to%
\begin{equation}
2\frac{\partial }{\partial r}C_{(k_{i})}+2(d-2\overline{k}%
_{i})C_{(k_{i})}+1+b_{1(k_{i})}C_{(k_{i})}+b_{2(k_{i})}C_{(k_{i})}^{2}=0.
\label{Cbb1}
\end{equation}%
It also has three classes of solutions, which are listed below.

1. For $b_{1(k_{i})}=b_{2(k_{i})}=0$,%
\begin{equation}
C_{(k_{i})}=\left\{
\begin{array}{c}
-\frac{1}{2(d-2\overline{k}_{i})}+\mathcal{O}(e^{-(d-2\overline{k}_{i})r}),\
d-2\overline{k}_{i}\neq 0 \\
-\frac{r}{2}+\mathcal{O}(1),\ d-2\overline{k}_{i}=0%
\end{array}%
\right. .  \label{Ck1}
\end{equation}

2. For $b_{1(k_{i})}\neq 0,\ b_{2(k_{i})}=0$,%
\begin{equation}
C_{(k_{i})}=\left\{
\begin{array}{c}
-\frac{1}{2(d-2\overline{k}_{i})+b_{1(k_{i})}}+\mathcal{O}(e^{-(d-2\bar{k}%
_{i}+b_{1(k_{i})}/2)r}),\ d-2\overline{k}_{i}\neq \frac{-b_{1(k_{i})}}{2} \\
-\frac{r}{2}+\mathcal{O}(1),\ d-2\overline{k}_{i}=\frac{-b_{1(k_{i})}}{2}%
\end{array}%
\right. .  \label{Ckb1}
\end{equation}

3. For $b_{1(k_{i})}=0$ and $b_{2(k_{i})}\neq 0$,

\begin{equation}
C_{(k_{i})}=\left\{
\begin{array}{c}
\frac{-1}{(d-2\overline{k}_{i})+\sqrt{(d-2\overline{k}_{i})^{2}-b_{2(k_{i})}}%
}+\mathcal{O}(e^{-\sqrt{(d-2\overline{k}_{i})^{2}-b_{2(k_{i})}}r}),\ d-2%
\overline{k}_{i}\neq \sqrt{b_{2(k_{i})}} \\
-\frac{1}{d-2\overline{k}_{i}}+\frac{2}{(d-2\overline{k}_{i})^{2}}\frac{1}{r}%
+\mathcal{O}(r^{-2}),\ d-2\overline{k}_{i}=\sqrt{b_{2(k_{i})}}%
\end{array}%
\right. .  \label{Ckb2}
\end{equation}%
Here we keep the coefficients up to the presence of integral constants.
Thus, the higher orders can be dropped since they should not be relevant to
any real divergent terms. As a result, we have obtained the coefficients at
the $i$-th order. Each coefficient is determined by four constants $d,%
\overline{k}_{i},b_{1(k_{i})},b_{2(k_{i})}$. It is worth pointing out that
there are two important special cases. One is relevant to the theories with
constant potential. Since $b_{1(k_{i})}=b_{2(k_{i})}=0$ and $k_{i}=2%
\overline{k}_{i}$ therein, the coefficients $C_{(k_{i})}$ are universal for
all counterterms at any given ($d,k_{i}$), see section C in Appendix. The
other is relevant to any logarithm divergence, which appears at $d-2%
\overline{k}_{i}=0$ or $-b_{1(k_{i})}/2$ and is associated with the
universal coefficient $-r/2$.

\section{General algorithm}

We will extract the spirit of the above benchmark and promote it to be a
more general algorithm, which can be stated as follows.

1. Derive the HJ equation and separate the counterterm part from it.

2. Take the formal expansion according to the divergence degree. Express any
order of counterterms as $U_{(k_{i})}=C_{(k_{i})}\left( r\right) \bar{U}%
_{(k_{i})}$ and build up the recursive equation%
\begin{equation}
2\frac{\partial }{\partial r}C_{(k_{i})}\bar{U}_{(k_{i})}+F(C_{(k_{i})})\bar{%
U}_{(k_{i})}+L_{A(k_{i})}=0.  \label{UCFL}
\end{equation}%
Here $F(C_{(k_{i})})$ is a function of the coefficient\footnote{%
For all the models in this paper, the function is the sum of linear and
quadratic terms. But it would be changed for a more general model.} but $%
L_{A(k_{i})}$ is independent with the coefficient.

3. Generate the exact ansatz of the counterterms iteratively from $%
L_{A(k_{i})}$.

4. Set $L_{A(k_{i})}=\bar{U}_{(k_{i})}=1$ in Eq. (\ref{UCFL}) to obtain an
ODE of the coefficient. Solve the ODE for a general solution.

Note that we have assumed that the solution of Eq. (\ref{UCFL}) without $%
L_{A(k_{i})}$ is not relevant to divergent counterterms\footnote{%
The counterexamples may be very rare, even if they would exist.}.

\section{Examples and assessments}

In section D of Appendix, we will use the above approach to work out the
counterterms explicitly in some typical holographic models. Keeping these
examples in mind, we will assess the approach from four aspects.

\subsection{Simplicity}

The main calculation we need is to take variation and solve ODE. The
difficulty of variation (with respect to the tensor) is greatly reduced by
some symbolic computing programs (such as the Mathematica package xAct \cite%
{xAct}). As for ODE, they are first order and usually simple, at least for
the holographic models we encounter in this paper. In particular, when the
general solution is found and the variation is prepared, the counterterms
can be obtained only by algebra calculation.

\subsection{Universality}

We have taken the model with the action (\ref{action}) as the benchmark.
This is a rather general holographic model. In fact, most models in Appendix
are its special cases. They include the theory of gravity coupled to a
scalar field with a general potential \cite{Papadimitriou2016}, the FGPW
model where one of two scalars has $\Delta =d/2$ \cite%
{Freedman9904,Elvang1603}, and the axion-dilaton model with constant
potential\footnote{%
In this model, the scalar fields are dual to marginal operators and they do
not enjoy the the same suppression as the scalar fields dual to relevant
operators. This issue is handled in \cite{Elvang1603} by allowing the
coefficients in the ansatz as the functions of marginal scalars. In our
work, it is not necessary. Different scalar fields are treated in the same
way.}. As the illustration of more general algorithm, we also study two
models that cannot be described by the action (\ref{action}). They are the
massive gravity that breaks the diffeomorphism symmetry \cite%
{Vegh1301,Davison2013,Blake2013,Blake2014,Cao1509} and the non-linear holographic axion
model \cite%
{Taylor1406,Baggioli2014,Li2108,Matteo2021}.

\subsection{Exactness}

The HJ approach to the HR of the axion-dilaton model has been studied in
\cite{Elvang1603}, where the first step is to postulate the ansatz for each
counterterm. For the highest order at $d=4$, their ansatz has 28 terms. Note
that the terms up to total derivatives have already been omitted.
Nevertheless, the redundancy is considerable, since the number of real
counterterms is only 16. The situation is more serious for the non-linear holographic
axion model: it is difficult to postulate the sufficient ansatz due to the non-integer powers. As a
comparison, we generate the exact ansatz in Appendix for all the models
including these two.

\subsection{Fluency}

As an advantage inherited from HJ approach \cite{Elvang1603,Wu1903}, the
logarithmic divergence does not cause any more trouble than the power
divergence. On the contrary, here it is somewhat simpler to deal with the
logarithmic divergence. This can be understood as follows. After generating
the exact ansatz at certain order with the logarithmic divergence, we
immediately know that all coefficients are $-r/2$, which has been pointed
out as the second special case below Eq. (\ref{Ckb2}). The logarithmic
divergence is present in general at the highest order when $d$ is even. It
also appears in massive gravity and holographic axion model when $d$ is odd.
Since the highest order has the most independent counterterms, the universal
coefficient brings considerable simplification.

\section{Summary}

In this work, we developed an alternative approach to holographic
renormalization based on the HJ formulation of gravity. Its distinguishing
feature is the generation of exact ansatz of counterterms. Although our
primary focus is on the technical aspects, this approach brings new
understandings of how counterterms are organized and generated. As one can
see, the $i$-th counterterm is determined by the exact ansatz $L_{A(k_{i})}$
and the coefficient function $F(C_{(k_{i})})$ in Eq. (\ref{UCFL}).
Meanwhile, the existence of general solutions to the coefficients informs us
all types of counterterms in a certain class of theory. In particular, the
benchmark model allows three types of counterterms, which are power law,
logarithmic and inverse logarithmic. Furthermore, the general conditions
under which these types appear may be helpful in designing a special UV
which is required by some bottom-up holographic models. For example, if we
expect a conformal anomaly when $d$ is odd, one way is to design a model
such that the counterterm seed has a term with half-integer inverse metrics.
This is exactly the fact that massive gravity exhibits. In the future, it
would be interesting to explore whether the current approach can be extended beyond the standard AdS/CFT,
which may break the conformal symmetry \cite{Ross0907,Baggio1107,Mann1107,Griffin1112,Chemissany1405}, keep the finite coupling \cite{Astefanesei0806,Liu0807,Kwon1106}, and even
deviate from the large N limit \cite{Skenderis2208}.

\section*{Acknowledgments}

We thank Matteo Baggioli, Xian-Hui Ge, Ioannis Papadimitriou, and Yu Tian
for helpful discussions. SFW was supported partially by NSFC grants
(No.11675097).\bigskip


\newpage \appendix*{Appendix} \setcounter{equation}{0} \renewcommand{%
\theequation}{A.\arabic{equation}}

\section{Decompose HJ equation}

Let's start from the HJ equation:%
\begin{equation}
H\left( \gamma _{ij},\Phi ^{I};\frac{\delta S_{\mathrm{os}}}{\delta \gamma
_{ij}},\frac{\delta S_{\mathrm{os}}}{\delta \Phi ^{I}}\right) +\frac{%
\partial S_{\mathrm{os}}}{\partial r}=0.  \label{HJ}
\end{equation}%
Suppose that $S_{\mathrm{ren}}$ and $S_{\mathrm{ct}}$ are the renormalized
part and the counterterm part of $S_{\mathrm{os}}$, respectively. Then Eq. (%
\ref{HJ}) can be decomposed into%
\begin{equation}
H_{\mathrm{ren}}+\frac{\partial S_{\mathrm{ren}}}{\partial r}-H_{\mathrm{ct}%
}-\frac{\partial S_{\mathrm{ct}}}{\partial r}=0.  \label{HJ2}
\end{equation}%
Here $H_{\mathrm{ct}}\equiv -(H-H_{\mathrm{ren}})$ and we define $H_{\mathrm{%
ren}}$ as the part of $H$ relevant to $S_{\mathrm{ren}}$. Using the
expression of Hamiltonian (\ref{H}), we read%
\begin{align}
H_{\mathrm{ren}}& =\int_{\partial M}d^{d}x\left[ 2\{-S_{\mathrm{ct}},S_{%
\mathrm{ren}}\}+\{S_{\mathrm{ren}},S_{\mathrm{ren}}\}\right] ,  \label{Hren}
\\
H_{\mathrm{ct}}& =-\int_{\partial M}d^{d}x\left[ \{S_{\mathrm{ct}},S_{%
\mathrm{ct}}\}+\sqrt{\gamma }L_{S}\right] ,
\end{align}%
where $L_{S}$ is the counterterm seed and the bracket $\{S_{\mathrm{a}},S_{%
\mathrm{b}}\}$ is defined through%
\begin{equation}
\{S_{\mathrm{a}},S_{\mathrm{b}}\}=\frac{1}{\sqrt{\gamma }}\left( \frac{%
\delta S_{\mathrm{a}}}{\delta \gamma _{ij}}\frac{\delta S_{\mathrm{b}}}{%
\delta \gamma _{kl}}\gamma _{ijkl}+\frac{1}{4}G^{IJ}\frac{\delta S_{\mathrm{a%
}}}{\delta \Phi ^{I}}\frac{\delta S_{\mathrm{b}}}{\delta \Phi ^{J}}\right) .
\end{equation}%
We can change Eq. (\ref{Hren}) a little as
\begin{equation}
H_{\mathrm{ren}}=\int_{\partial M}d^{d}x\left[ 2\{S_{\mathrm{os}},S_{\mathrm{%
ren}}\}-\{S_{\mathrm{ren}},S_{\mathrm{ren}}\}\right] .  \label{Hren1}
\end{equation}%
Note that the second term in Eq. (\ref{Hren1}) is vanishing near the
boundary. This is because it is much smaller than the first term by
definition and we will show below that the first term is not greater than $%
\mathcal{O}(1)$, see Eq. (\ref{Hren2}).

To calculate the variations of $S_{\mathrm{os}}$, we can equate two forms of
the momentum%
\begin{align}
\frac{\delta S_{\mathrm{os}}}{\delta \gamma _{ij}}& =\frac{\partial L}{%
\partial \dot{\gamma}_{ij}}=\sqrt{\gamma }(\mathcal{K}^{ij}-\mathcal{K}%
\gamma ^{ij}),  \label{twomomenta1} \\
\frac{\delta S_{\mathrm{os}}}{\delta \Phi ^{I}}& =\frac{\partial L}{\partial
\dot{\Phi}^{I}}=2\sqrt{\gamma }G_{IJ}\dot{\Phi}^{J},  \label{twomomenta2}
\end{align}%
where the extrinsic curvature tensor $\mathcal{K}_{\ j}^{i}=\gamma ^{ik}\dot{%
\gamma}_{kj}/2$. Inserting Eqs. (\ref{twomomenta1}-\ref{twomomenta2}) into
Eq. (\ref{Hren1}), we find%
\begin{equation}
H_{\mathrm{ren}}=\int_{\partial M}d^{d}x\left( \frac{\delta S_{\mathrm{ren}}%
}{\delta \gamma _{ij}}\dot{\gamma}_{ij}+\frac{\delta S_{\mathrm{ren}}}{%
\delta \Phi ^{I}}\dot{\Phi}^{I}\right) .  \label{Hren2}
\end{equation}%
Furthermore, we consider that $S_{\mathrm{ren}}$ can be taken as the
functional of ($\bar{\gamma}_{ij},\bar{\Phi}^{I}$) or ($\gamma _{ij},\Phi
^{I},r$). This is viewed from the field theory and its gravity dual,
respectively. Keeping this in mind, we can derive%
\begin{eqnarray}
\frac{\partial S_{\mathrm{ren}}}{\partial r}+H_{\mathrm{ren}} &=&\frac{%
\partial S_{\mathrm{ren}}\left( \gamma _{ij},\Phi ^{I},r\right) }{\partial r}%
+\int_{\partial M}d^{d}x\left( \frac{\delta S_{\mathrm{ren}}}{\delta \gamma
_{ij}}\dot{\gamma}_{ij}+\frac{\delta S_{\mathrm{ren}}}{\delta \Phi ^{I}}\dot{%
\Phi}^{I}\right) \\
&=&\frac{dS_{\mathrm{ren}}(\bar{\gamma}_{ij},\bar{\Phi}^{I})}{dr}=0.
\end{eqnarray}%
Combining it with Eq. (\ref{HJ2}), we can obtain the counterterm part of the
HJ equation\footnote{%
This equation has been derived before in \cite{Wu1903}. Here the procedure
is more general since we have not invoked the explicit asymptotic behavior
of the metric and scalar fields. Moreover, Eq. (\ref{CPHJ}) is similar to
Eq. (27) in \cite{Papadimitriou2016}, and their relation has been explained
in \cite{Wu1903}.}%
\begin{equation}
H_{\mathrm{ct}}+\frac{\partial S_{\mathrm{ct}}}{\partial r}=0.  \label{CPHJ}
\end{equation}

\section{Variation identities}

We will study two variation identities. Suppose that a boundary invariant
with divergence degree $k$ can be formally written as%
\begin{equation}
U_{(k)}=\left( \gamma ^{ab}\cdots \right) \left( \Phi ^{I}\cdots \right)
\left( R_{\ bcd}^{a}\cdots \right) \left( \nabla _{a}Y_{b\cdots }\cdots
\right) .  \label{generalU}
\end{equation}%
Here $\left( X\cdots \right) $ denotes the product of (one or more) $X$ and $%
Y_{b\cdots }$ denotes any tensor which is made by the product of $\Phi ^{I}$
and $R_{\ bcd}^{a}$ as well as their covariant derivatives.

For convenience, we define an operator%
\begin{equation}
\int X\bar{\delta}\equiv \gamma ^{ij}\frac{1}{\sqrt{\gamma }}\int d^{d}x%
\sqrt{\gamma }X\frac{\delta }{\delta \gamma ^{ij}},
\end{equation}%
where $X$ can denote any tensor. Using the chain rule, we have%
\begin{eqnarray}
\int \bar{\delta}U_{(k)} &=&\int \bar{\delta}\left( \gamma ^{ab}\cdots
\right) \left( \Phi ^{I}\cdots \right) \left( R_{\ bcd}^{a}\cdots \right)
\left( \nabla _{a}Y_{b\cdots }\cdots \right)  \notag \\
&&+\int \left( \gamma ^{ab}\cdots \right) \bar{\delta}\left( \Phi ^{I}\cdots
\right) \left( R_{\ bcd}^{a}\cdots \right) \left( \nabla _{a}Y_{b\cdots
}\cdots \right)  \notag \\
&&+\int \left( \gamma ^{ab}\cdots \right) \left( \Phi ^{I}\cdots \right)
\bar{\delta}\left( R_{\ bcd}^{a}\cdots \right) \left( \nabla _{a}Y_{b\cdots
}\cdots \right)  \notag \\
&&+\int \left( \gamma ^{ab}\cdots \right) \left( \Phi ^{I}\cdots \right)
\left( R_{\ bcd}^{a}\cdots \right) \bar{\delta}\left( \nabla _{a}Y_{b\cdots
}\cdots \right) .
\end{eqnarray}%
We will calculate the four terms respectively:%
\begin{eqnarray}
\int \bar{\delta}\left( \gamma ^{ab}\cdots \right) \left( \Phi ^{I}\cdots
\right) \left( R_{\ bcd}^{a}\cdots \right) \left( \nabla _{a}Y_{b\cdots
}\cdots \right) &=&\bar{k}U_{(k)},  \label{dU1} \\
\int \left( \gamma ^{ab}\cdots \right) \bar{\delta}\left( \Phi ^{I}\cdots
\right) \left( R_{\ bcd}^{a}\cdots \right) \left( \nabla _{a}Y_{b\cdots
}\cdots \right) &=&0,  \label{dU2} \\
\int \left( \gamma ^{ab}\cdots \right) \left( \Phi ^{I}\cdots \right) \bar{%
\delta}\left( R_{\ bcd}^{a}\cdots \right) \left( \nabla _{a}Y_{b\cdots
}\cdots \right) &=&\mathrm{TD,}  \label{dU3} \\
\int \left( \gamma ^{ab}\cdots \right) \left( \Phi ^{I}\cdots \right) \left(
R_{\ bcd}^{a}\cdots \right) \bar{\delta}\left( \nabla _{a}Y_{b\cdots }\cdots
\right) &=&\mathrm{TD.}  \label{dU4}
\end{eqnarray}%
The former two equations are obvious. Note that $\bar{k}$ is the number of
inverse metric in $\left( \gamma ^{ab}\cdots \right) $. The latter two
equations are derived as follows:%
\begin{eqnarray}
&&\int X\bar{\delta}R_{\;bcd}^{a}  \notag \\
&=&\int X\left( \nabla _{c}\bar{\delta}\Gamma _{bd}^{a}-\nabla _{d}\bar{%
\delta}\Gamma _{bc}^{a}\right)  \notag \\
&=&\int \left( -\nabla _{c}X\bar{\delta}\Gamma _{bd}^{a}+\nabla _{d}X\bar{%
\delta}\Gamma _{bc}^{a}\right)  \notag \\
&=&\int -\nabla _{c}X\frac{1}{2}\gamma ^{ae}\left( \nabla _{b}\bar{\delta}%
\gamma _{ed}+\nabla _{d}\bar{\delta}\gamma _{be}-\nabla _{e}\bar{\delta}%
\gamma _{bd}\right) -\left( c\rightarrow d\right)  \notag \\
&=&\int \frac{1}{2}\gamma ^{ae}\left( \nabla _{b}\nabla _{c}X\bar{\delta}%
\gamma _{ed}+\nabla _{d}\nabla _{c}X\bar{\delta}\gamma _{be}-\nabla
_{e}\nabla _{c}X\bar{\delta}\gamma _{bd}\right) -\left( c\rightarrow d\right)
\notag \\
&=&-\frac{1}{2}\left( \nabla _{b}\nabla _{c}X\delta _{d}^{a}+\nabla
_{d}\nabla _{c}X\delta _{b}^{a}-\nabla ^{a}\nabla _{c}X\gamma _{bd}\right)
-\left( c\rightarrow d\right) ,  \label{XDR}
\end{eqnarray}%
\begin{eqnarray}
&&\int X\bar{\delta}\left( \nabla _{i}Y_{j...}\right)  \notag \\
&=&\int X\bar{\delta}\left( \partial _{i}Y_{j...}-\Gamma
_{ij}^{k}Y_{k...}+\cdots \right)  \notag \\
&=&\int X\left[ \left( -\bar{\delta}\Gamma _{ij}^{k}Y_{k...}+\cdots \right)
+\nabla _{i}\bar{\delta}Y_{j...}\right]  \notag \\
&=&\int \left[ -X\frac{1}{2}\gamma ^{ke}\left( \nabla _{i}\bar{\delta}\gamma
_{ej}+\nabla _{j}\bar{\delta}\gamma _{ie}-\nabla _{e}\bar{\delta}\gamma
_{ij}\right) Y_{k...}+\cdots +X\nabla _{i}\bar{\delta}Y_{j...}\right]  \notag
\\
&=&\int \left\{ \frac{1}{2}\left[ \nabla _{i}\left( XY_{...}^{e}\right) \bar{%
\delta}\gamma _{ej}+\nabla _{j}\left( XY_{...}^{e}\right) \bar{\delta}\gamma
_{ie}-\nabla _{e}\left( XY_{...}^{e}\right) \bar{\delta}\gamma _{ij}\right]
+\cdots -\nabla _{i}X\bar{\delta}Y_{j...}\right\}  \notag \\
&=&\frac{1}{2}\left[ -\nabla _{i}\left( XY_{j...}\right) -\nabla _{j}\left(
XY_{i...}\right) +\nabla _{k}\left( XY_{...}^{k}\right) \gamma _{ij}\right]
+\cdots -\int \nabla _{i}X\bar{\delta}Y_{j...}  \label{XDY}
\end{eqnarray}%
Here $\cdots $ in Eq. (\ref{XDY}) denote the contributions according to the
suppressed index $_{...}$ in the tensor $Y_{j...}$. Moreover, we note that
the last term above is a total derivative, which can be seen by iteratively
using Eq. (\ref{dU2}), Eq. (\ref{dU3}), and Eq. (\ref{XDY}). Combining Eqs. (%
\ref{dU1})-(\ref{dU4}), we have proved\footnote{%
This identity may still hold even if the counterterm is more general than
Eq. (\ref{generalU}). For examples, see the massive gravity in section D and
the holographic axion model in section E.}%
\begin{equation}
\int \bar{\delta}U_{(k)}=\bar{k}U+\mathrm{TD.}  \label{dbarU}
\end{equation}%
From the form of Eq. (\ref{generalU}), it is obvious that there is a
constraint $2\overline{k}\leq k$.

Next, we turn to the operator%
\begin{equation}
\int X\tilde{\delta}\equiv \Phi ^{I}\frac{1}{\sqrt{\gamma }}\int d^{d}x\sqrt{%
\gamma }X\frac{\delta }{\delta \Phi ^{I}}.
\end{equation}%
Acting it on Eq. (\ref{generalU}), we read%
\begin{eqnarray}
\int \tilde{\delta}U_{(k)} &=&\int \tilde{\delta}\left( \gamma ^{ab}\cdots
\right) \left( \Phi ^{I}\cdots \right) \left( R_{\ bcd}^{a}\cdots \right)
\left( \nabla _{a}Y_{b\cdots }\cdots \right)  \notag \\
&&+\int \left( \gamma ^{ab}\cdots \right) \tilde{\delta}\left( \Phi
^{I}\cdots \right) \left( R_{\ bcd}^{a}\cdots \right) \left( \nabla
_{a}Y_{b\cdots }\cdots \right)  \notag \\
&&+\int \left( \gamma ^{ab}\cdots \right) \left( \Phi ^{I}\cdots \right)
\tilde{\delta}\left( R_{\ bcd}^{a}\cdots \right) \left( \nabla
_{a}Y_{b\cdots }\cdots \right)  \notag \\
&&+\int \left( \gamma ^{ab}\cdots \right) \left( \Phi ^{I}\cdots \right)
\left( R_{\ bcd}^{a}\cdots \right) \tilde{\delta}\left( \nabla
_{a}Y_{b\cdots }\cdots \right) .  \label{dU44}
\end{eqnarray}%
Each line yields%
\begin{eqnarray}
\int \tilde{\delta}\left( \gamma ^{ab}\cdots \right) \left( \Phi ^{I}\cdots
\right) \left( R_{\ bcd}^{a}\cdots \right) \left( \nabla _{a}Y_{b\cdots
}\cdots \right) &=&0,  \label{dU5} \\
\int \left( \gamma ^{ab}\cdots \right) \tilde{\delta}\left( \Phi ^{I}\cdots
\right) \left( R_{\ bcd}^{a}\cdots \right) \left( \nabla _{a}Y_{b\cdots
}\cdots \right) &=&\tilde{k}_{1}U_{(k)},  \label{dU6} \\
\int \left( \gamma ^{ab}\cdots \right) \left( \Phi ^{I}\cdots \right) \tilde{%
\delta}\left( R_{\ bcd}^{a}\cdots \right) \left( \nabla _{a}Y_{b\cdots
}\cdots \right) &=&\mathrm{0,}  \label{dU7} \\
\int \left( \gamma ^{ab}\cdots \right) \left( \Phi ^{I}\cdots \right) \left(
R_{\ bcd}^{a}\cdots \right) \tilde{\delta}\left( \nabla _{a}Y_{b\cdots
}\cdots \right) &=&\tilde{k}_{2}U_{(k)}\mathrm{,}  \label{dU8}
\end{eqnarray}%
where $\tilde{k}_{1}$ and $\tilde{k}_{2}$ denote the numbers of $\Phi ^{I}$
in $\left( \Phi ^{I}\cdots \right) $ and $\left( \nabla _{a}Y_{b\cdots
}\cdots \right) $, respectively. Note that one can use%
\begin{equation}
\int X\tilde{\delta}\left( \nabla _{a}Y_{b\cdots }\right) =\int -\nabla _{a}X%
\tilde{\delta}Y_{b\cdots }
\end{equation}%
and Eqs. (\ref{dU6})-(\ref{dU7}) iteratively to prove Eq. (\ref{dU8}). Thus,
we have%
\begin{equation}
\int \tilde{\delta}U_{(k)}=\tilde{k}U_{(k)},  \label{dtU}
\end{equation}%
where $\tilde{k}=\tilde{k}_{1}+\tilde{k}_{2}$ means the number of $\Phi ^{I}$
in $U_{(k)}$. Note that there is also a constraint $2\overline{k}+\Delta _{-}%
\tilde{k}\leq k$ associated with Eq. (\ref{dtU}). Here we have set $\Phi
^{I}\sim e^{-\Delta _{-}r}$.

\section{Mass parameters}

The two parameters $b_{(1)}$ and $b_{(2)}$ are defined by%
\begin{eqnarray}
L_{P(k_{i})} &=&\sum\limits_{\substack{ m,n\rightarrow i  \\ m=i\ or\ n=i}}%
G^{IJ}P_{I(k_{m})}P_{J(k_{n})}+\sum\limits_{\substack{ m,n\rightarrow i  \\ %
m=i\ and\ n=i}}G^{IJ}P_{I(k_{m})}P_{J(k_{n})}  \label{LPk2} \\
&=&C_{(k_{i})}\bar{U}_{(k_{i})}b_{1(k_{i})}+C_{(k_{i})}^{2}\bar{U}%
_{(k_{i})}b_{2(k_{i})},
\end{eqnarray}%
where%
\begin{equation}
P_{I(k_{m})}=\int \frac{\delta U_{(k_{m})}}{\delta \Phi ^{I}},\
P_{J(k_{n})}=\int \frac{\delta U_{(k_{n})}}{\delta \Phi ^{J}}.
\end{equation}%
Now let's calculate the divergence degrees of two terms in Eq. (\ref{LPk2}).
Without loss of generality, we have%
\begin{eqnarray}
\left[ G^{IJ}P_{I(k_{i})}P_{J(k_{n})}\right] &=&\left[ G^{IJ}\right] +\left[
P_{I(k_{i})}\right] +\left[ P_{J(k_{n})}\right]  \notag \\
&=&\left[ G^{IJ}\right] +k_{i}+k_{n}-\left[ \Phi ^{I}\right] -\left[ \Phi
^{J}\right] =k_{i},  \label{GPP1}
\end{eqnarray}%
\begin{eqnarray}
\left[ G^{IJ}P_{I(k_{i})}P_{J(k_{i})}\right] &=&\left[ G^{IJ}\right] +\left[
P_{I(k_{i})}\right] +\left[ P_{J(k_{i})}\right]  \notag \\
&=&\left[ G^{IJ}\right] +2k_{i}-\left[ \Phi ^{I}\right] -\left[ \Phi ^{J}%
\right] =k_{i},  \label{GPP2}
\end{eqnarray}%
where $\left[ \cdots \right] $ denotes the divergence degree of $\cdots $.
We will analyze the parameters $b_{(1)}$ and $b_{(2)}$ in terms of Eq. (\ref%
{GPP1}) and Eq. (\ref{GPP2}). We will focus on the situations related to
holographic models studied in next section.

1. Since $k_{n}\neq k_{i}$ in Eq. (\ref{GPP1}), the two equations cannot
hold at the same time. This means either $b_{1(k_{i})}=0$ or $b_{2(k_{i})}=0$%
.

2. Consider the benchmark model with constant potential, where $\left[ \Phi
^{I}\right] =\left[ \Phi ^{J}\right] =0$ and $\left[ G^{IJ}\right] =0$. Then
neither Eq. (\ref{GPP1}) nor Eq. (\ref{GPP2}) holds and we have $%
b_{1(k_{i})}=b_{2(k_{i})}=0$ and $k_{i}=2\bar{k}_{i}$.

3. Consider the benchmark model with massive scalar fields and suppose $%
G^{IJ}$ as a diagonal constant matrix. The two equations have the solutions $%
k_{n}=2[\Phi ^{I}]$ and $k_{i}=2[\Phi ^{I}]$. Keeping this in mind, one can
find%
\begin{equation}
b_{1(k_{i})}=\frac{4G^{II}V_{II}\tilde{k}_{i}}{d+\sqrt{d^{2}-b_{2(2\left[
\Phi ^{I}\right] )}}},\ b_{2(2\left[ \Phi ^{I}\right] )}=-4G^{II}V_{II},
\label{b12}
\end{equation}%
where $\tilde{k}_{i}$ is the number of scalar field $\Phi ^{I}$ in $%
U_{(k_{i})}$, $V_{II}$ is the constant before\ the term $\Phi ^{I}\Phi ^{I}$
in the potential, $G^{II}$ is the inverse of scalar couplings, and the same
index $I$ is not summed up. Note that we have used Eq. (\ref{dtU}) to derive
Eq. (\ref{b12}). Since $G^{II}V_{II}=m^{2}$ and $\Delta _{-}=\frac{d}{2}-%
\sqrt{\left( \frac{d}{2}\right) ^{2}+m^{2}}$, where $m$ is the mass of $\Phi
^{I}$, Eq. (\ref{b12}) can be reduced to%
\begin{equation}
b_{1(k_{i})}=-2\tilde{k}_{i}\Delta _{-},\ b_{2(2\Delta _{-})}=-4m^{2}.
\label{b122}
\end{equation}%
We therefore refer $b_{(1)}$ and $b_{(2)}$ as the mass parameters.

4. In section D.5, we will study the holographic axion model, where the
divergence degree $\left[ \Phi^{I}\right] =0$ but the effective couplings $%
\left[ \bar{G}^{IJ}\right] \neq 0$. Apparently, two equations have the
solutions $k_{n}=-\left[ \bar{G}^{IJ}\right] $ and $k_{i}=-\left[ \bar{G}%
^{IJ}\right] $. However, they are not real solutions since $k_{i}>-\left[
\bar{G}^{IJ}\right] $ therein. So we still have $b_{1(k_{i})}=b_{2(k_{i})}=0$%
.

\section{Applications}

We will apply the HJ approach developed in main text to five typical models.
The HR of the former four models has been studied before by different
approaches \cite{Papadimitriou1106,Papadimitriou2016,Elvang1603,Wu1903}. The
two special cases of the last model have also been studied in \cite%
{Andrade1311} and \cite{Taylor1406}. One can see that our results are
consistent with theirs.

\subsection{General potential}

Suppose that the boundary dimension is $d=4$ and there is only one scalar
field $\phi $. From HJ equation, we read the (inverse) coupling $G^{11}=2$
and the counterterm seed%
\begin{equation}
L_{S}=R-\frac{1}{2}\gamma ^{ij}\partial _{i}\phi \partial _{j}\phi -V\left(
\phi \right) ,
\end{equation}%
where the scalar potential is
\begin{equation}
V\left( \phi \right) =V_{0}+V_{1}\phi +V_{2}\phi ^{2}+V_{3}\phi
^{3}+V_{4}\phi ^{4}+\cdots ,
\end{equation}%
with $V_{0}=12$, $V_{1}=0$, and $V_{2}=-3/2$. The mass square is $m^{2}=-3$
and the scaling dimension is $\Delta =3$.

We will derive the counterterms by three steps.

1. Generate ansatz

With $U_{(k_{0})}=-3$ and $L_{S}$ in hand, we can calculate Eq. (\ref{LAG})
iteratively and generate the ansatz at all order with $k_{i}\leq d$:%
\begin{equation}
L_{A(2)}=R-V_{2}\phi ^{2},\ L_{A(3)}=-V_{3}\phi ^{3},  \label{LA1}
\end{equation}%
\begin{equation}
L_{A(4)}=-\frac{1}{2}\gamma ^{ij}\partial _{i}\phi \partial _{j}\phi
-V_{4}\phi ^{4}+4Y_{(2)ij}Y_{(2)}^{ij}+2P_{(3)}^{2}-U_{(2)}^{2}-\frac{1}{3}%
\left( U_{(2)}-2Y_{(2)}\right) ^{2}.  \label{LA41}
\end{equation}

2. Take variation

What we need is%
\begin{equation}
Y_{(2)ij}=C_{(2)}^{(1)}R_{ij},\ P_{(3)}=-3V_{3}\phi ^{2}C_{(3)}.  \label{YP1}
\end{equation}%
Note that the upper index $(1)$ in $C_{(2)}^{(1)}$ is used to indicate that $%
C_{(2)}^{(1)}$ is associated with the first term in $L_{A(2)}$.

3. Specify coefficients

Using Eq. (\ref{b122}), one can read the mass parameters%
\begin{equation}
b_{2(2)}^{(2)}=12,\ b_{1(3)}=-6.
\end{equation}%
The coefficients can be obtained by Eqs. (\ref{Ck1}), (\ref{Ckb1}) and (\ref%
{Ckb2}). In particular, keep in mind that they are universal at the highest
order: $C_{(d)}=-\frac{r}{2}$. Then we have%
\begin{equation}
C_{(2)}^{(1)}=-\frac{1}{4},\ C_{(2)}^{(2)}=-\frac{1}{6},\ C_{(3)}=-\frac{1}{2%
},\ C_{(4)}=-\frac{r}{2}.  \label{C1}
\end{equation}

We collect $U_{(k_{i})}$ at all orders (except $U_{(k_{0})}=1-d$):%
\begin{equation}
U_{(2)}=-\frac{1}{4}\left( R+\phi ^{2}\right) ,\ U_{(3)}=\frac{1}{2}%
V_{3}\phi ^{3},\ U_{(4)}=-\frac{r}{2}L_{A(4)}.  \label{U23}
\end{equation}%
Putting Eq. (\ref{LA41}), Eq. (\ref{YP1}), Eq. (\ref{C1}), and Eq. (\ref{U23}%
) together, it yields%
\begin{equation}
U_{(4)}=-\frac{r}{2}\left[ \frac{1}{4}R_{ij}R^{ij}-\frac{1}{12}R^{2}-\frac{1%
}{2}\gamma ^{ij}\partial _{i}\phi \partial _{j}\phi -\frac{1}{12}R\phi ^{2}+(%
\frac{9}{2}V_{3}^{2}-V_{4}-\frac{1}{12})\phi ^{4}\right] .
\end{equation}

\subsection{Dimension $\Delta =d/2$}

Suppose that the boundary dimension is $d=4$ and there are two scalar fields
$\psi $ and $\phi $. The couplings are the identity matrix $G^{IJ}=\mathrm{%
diag}(1,1)$ and the counterterm seed is%
\begin{equation}
L_{S}=R-\gamma ^{ij}\partial _{i}\psi \partial _{j}\psi -\gamma
^{ij}\partial _{i}\phi \partial _{j}\phi -V\left( \psi ,\phi \right) ,
\end{equation}%
where the scalar potential is%
\begin{equation}
V\left( \psi ,\phi \right) =-12-4\phi ^{2}-3\psi ^{2}+\psi ^{4}+\cdots .
\end{equation}%
The mass squares are $m^{2}=-3$ and \thinspace $-4$ for two scalars fields,
respectively. Accordingly, the scaling dimensions are $\Delta =3$ and $2$.
The latter indicates a nontrivial asymptotic behavior $\phi \sim re^{-2r}$.
In order to sort the counterterms that involve $\phi $, we set $\phi \sim
e^{-\bar{\Delta}r}$ where $\bar{\Delta}=2-\ln r/r$ is not a number. However,
it is convenient to understand $\bar{\Delta}$ as an effective dimension with
$1<\bar{\Delta}<2$.

In the following, we will derive the counterterms. Since the procedure is
similar to above example, we will omit unnecessary text descriptions for
brevity.

1. Generate ansatz%
\begin{eqnarray}
L_{A(2)} &=&R+3\psi ^{2},\ L_{A(2\bar{\Delta})}=4\phi ^{2}, \\
L_{A(4)} &=&-\gamma ^{ij}\partial _{i}\psi \partial _{j}\psi -\psi
^{4}+4Y_{(2)ij}Y_{(2)}^{ij}-U_{(2)}^{2}-\frac{1}{3}\left(
U_{(2)}-2Y_{(2)}\right) ^{2}.
\end{eqnarray}
Note that the divergence degree of $L_{A(2\bar{\Delta})}$ is $2\bar{\Delta}$%
, which is not a number. But this did not cause any problems with our approach.

2. Take variation%
\begin{equation}
Y_{(2)ij}=C_{(2)}^{(1)}R_{ij}.
\end{equation}

3. Specify coefficients%
\begin{equation}
b_{2(2)}^{(2)}=12,\ b_{2(2\bar{\Delta})}=16.
\end{equation}%
\begin{equation}
C_{(2)}^{(1)}=-\frac{1}{4},\ C_{(2)}^{(2)}=-\frac{1}{6},\ C_{(2\bar{\Delta}%
)}=-\frac{1}{4}+\frac{1}{8r},\ C_{(4)}=-\frac{r}{2}.
\end{equation}

We collect $U_{(k_{i})}$ at all orders:%
\begin{equation}
U_{(2)}=-\frac{1}{4}\left( R+2\psi ^{2}\right) ,\ U_{(2\bar{\Delta})}=\left(
-1+\frac{1}{2r}\right) \phi ^{2},\ U_{(4)}=-\frac{r}{2}L_{A(4)}.
\end{equation}%
One can find that $U_{(4)}$ can be rewritten as%
\begin{equation}
U_{(4)}=-\frac{r}{2}\left( \frac{1}{4}R_{ij}R^{ij}-\frac{1}{12}R^{2}-\gamma
^{ij}\partial _{i}\psi \partial _{j}\psi -\frac{1}{6}R\psi ^{2}-\frac{4}{3}%
\psi ^{4}\right) .
\end{equation}

\subsection{Constant potential}

Suppose that the boundary dimension is $d=4$ and there are two massless
scalar fields: the dilaton $\phi $ and the axion $\chi $. The couplings are $%
G^{IJ}=\mathrm{diag}(1,1/Z)$. Here $Z$ is a function of $\phi $ and $I=1,2$
refer $\phi $ and $\chi $, respectively. The counterterm seed is%
\begin{equation}
L_{S}=R-\gamma ^{ij}\nabla _{i}\phi \nabla _{j}\phi -Z\gamma ^{ij}\nabla
_{i}\chi \nabla _{j}\chi -V.
\end{equation}%
Unlike the previous two examples, now the potential is constant $V=-12$.

1. Generate ansatz%
\begin{eqnarray}
L_{A(2)} &=&R-\gamma ^{ij}\nabla _{i}\phi \nabla _{j}\phi -Z\gamma
^{ij}\nabla _{i}\chi \nabla _{j}\chi ,  \label{LA2} \\
L_{A(4)} &=&P_{1(2)}^{2}+\frac{1}{Z}P_{2(2)}^{2}+4Y_{(2)ij}Y_{(2)}^{ij}-%
\frac{4}{3}U_{(2)}^{2}.  \label{LA4}
\end{eqnarray}

2. Take variation%
\begin{eqnarray}
Y_{(2)ij} &=&C_{(2)}\left( R_{ij}-\nabla _{i}\phi \nabla _{j}\phi -Z\nabla
_{i}\chi \nabla _{j}\chi \right) , \\
P_{1(2)} &=&C_{(2)}\left( 2\nabla ^{2}\phi -Z^{\prime }\gamma ^{ij}\nabla
_{i}\chi \nabla _{j}\chi \right) ,\ P_{2(2)}=2C_{(2)}\gamma ^{ij}\nabla
_{i}(Z\nabla _{j}\chi ).
\end{eqnarray}

3. Specify coefficients

Since the potential is constant, we have $b_{1(k_{i})}=b_{2(k_{i})}=0$ and $2%
\overline{k}_{i}=k_{i}$. This indicates $C_{(k_{i})}=-\frac{1}{2(d-k_{i})}$\
for $d\neq k_{i}$ or $-\frac{r}{2}$ for $d=k_{i}$. Concretely, we have%
\begin{equation}
C_{(2)}=-\frac{1}{4},\ C_{(4)}=-\frac{r}{2}\text{.}
\end{equation}

We collect $U_{(k_{i})}$ at all orders:%
\begin{equation}
U_{(2)}=-\frac{1}{4}L_{A(2)},\ U_{(4)}=-\frac{r}{2}L_{A(4)}.
\end{equation}%
Note that $U_{(4)}$ can be expanded as 16 independent terms \cite{Elvang1603}%
.

\subsection{Massive gravity}

Massive gravity cannot be described by the benchmark action in the main
text. Particularly, it breaks the diffeomorphism symmetry, which indicates
that the ADM metric cannot be gauged as usual. Nevertheless, it is found
that for the popular holographic model \cite%
{Vegh1301,Davison2013,Blake2013,Blake2014,Cao1509}, the lapse still can be
normalized and sometimes the shift vector is falling off fast enough near
the boundary. Thus, the HJ equation and its counterterm part still have the
same form as before, except that the counterterm seed is changed as \cite%
{Wu1903}%
\begin{equation}
L_{S}=R+d(d-1)+\sum\limits_{n=1}^{4}\alpha _{n}e_{n}(X),
\end{equation}%
where $e_{n}(X)$ are the symmetric polynomials of the eigenvalues of the $%
d\times d$ matrix $X{^{i}}_{j}$,%
\begin{eqnarray}
e_{1} &=&[X],\quad e_{2}=[X]^{2}-[X^{2}],\quad
e_{3}=[X]^{3}-3[X][X^{2}]+2[X^{3}], \\
e_{4} &=&[X]^{4}-6[X]^{2}[X]+8[X^{3}][X]+3[X^{2}]^{2}-6[X^{4}].
\end{eqnarray}%
Note that we have denoted $[X]=${$X$}${^{i}}_{i}$ and will set $d=4$.
Obviously, the matrix {$X$}${^{i}}_{j}$ is the key object, which is defined
as the square root of $\gamma ^{ij}f_{ij}$. Here $f_{ij}$ are the reference
metric on the boundary.

The HR of massive gravity can be performed by the general algorithm in main
text. One can expect that it is similar to the last model. This is because
their coefficient functions $F(C_{(k_{i})})$ in the recursive equations (\ref%
{UCFL}) are equal to the same $2(d-2\overline{k}_{i})$. Note that $e_{n}(X)$
in the counterterm seed and other ansatz generated below do not take the
form as Eq. (\ref{generalU}) but the variation identity (\ref{dbarU}) still
holds, where the number $\bar{k}$ is counted by treating {$X$}${^{i}}_{j}$
as the object with 1/2 inverse metric \cite{Wu1903}.

1. Generate ansatz%
\begin{eqnarray}
L_{A(1)} &=&\alpha _{1}e_{1}, \\
L_{A(2)} &=&\alpha _{2}e_{2}+R+4Y_{(1)ij}Y_{(1)}^{ij}-U_{(1)}^{2}, \\
L_{A(3)} &=&\alpha _{3}e_{3}+8Y_{(1)ij}Y_{(2)}^{ij}-2U_{(1)}U_{(2)}, \\
L_{A(4)} &=&\alpha _{4}e_{4}+4Y_{(2)ij}Y_{(2)}^{ij}+8Y_{(1)ij}Y_{(3)}^{ij}-%
\frac{4}{3}U_{(2)}^{2}-2U_{(1)}U_{(3)}.
\end{eqnarray}

2. Take variation%
\begin{eqnarray}
Y_{(1)ij} &=&C_{(1)}\alpha _{1}\frac{1}{2}X_{ij},  \label{Y1ij} \\
Y_{(2)ij} &=&C_{(2)}\left[ R_{ij}+\left( \alpha _{2}-C_{(1)}^{2}\alpha
_{1}^{2}\right) \left( 2[X]X_{ij}-[X^{2}]_{ij}\right) \right] ,  \label{Y2ij}
\end{eqnarray}%
\begin{eqnarray}
Y_{(3)ij} &=&C_{(3)}\Big\{\left( \frac{\alpha _{1}^{3}}{144}-\frac{\alpha
_{1}\alpha _{2}}{4}+3\alpha _{3}\right) \left( \frac{1}{2}[X]^{2}X_{ij}-%
\frac{1}{2}X_{ij}[X^{2}]-[X][X^{2}]_{ij}+[X^{3}]_{ij}\right)  \notag \\
&&+\frac{\alpha _{1}}{12}\left( (2\nabla ^{k}\nabla _{(j}X_{i)k}-\nabla
^{k}\nabla _{k}X_{ij}-\nabla _{i}\nabla _{j}[X])-3{R}_{k(i}{X_{j)}^{k}}%
\right)  \notag \\
&&+\frac{\alpha _{1}}{12}\left( [X]R_{ij}+\frac{1}{2}RX_{ij}+\gamma
_{ij}\nabla ^{k}\nabla _{k}X-\nabla _{i}\nabla _{j}[X]\right) \Big\}.
\end{eqnarray}%
Here we have entered $C_{(1)}$ and $C_{(2)}$ given below to simplify $%
Y_{(3)ij}$.

3. Specify coefficients%
\begin{equation}
C_{(1)}=-\frac{1}{6},\ C_{(2)}=-\frac{1}{4},\ C_{(3)}=-\frac{1}{2},\
C_{(4)}=-\frac{r}{2}.  \label{C14}
\end{equation}

We collect $U_{(k_{i})}$ at all orders:%
\begin{equation}
U_{(1)}=-\frac{1}{6}L_{A(1)},\ U_{(2)}=-\frac{1}{4}L_{A(2)},\ U_{(3)}=-\frac{%
1}{2}L_{A(3)},\ U_{(4)}=-\frac{r}{2}L_{A(4)}.
\end{equation}

\subsection{Non-integer power}

We will study an interesting model in AdS/CMT: the holographic axion model
\cite{Matteo2021}. We write down its action%
\begin{equation}
S=-\int_{M}d^{4}x\sqrt{g}\left( \mathcal{R}+6-V(X)\right) .
\end{equation}%
It has the derivative term $V=X^{n}$, where $X=g^{\mu \nu }X_{\mu \nu }$ and $%
X_{\mu \nu }=G_{IJ}\partial _{\mu }\chi ^{I}\partial _{\nu }\chi ^{J}$ with $%
G_{IJ}=\frac{1}{2}\mathrm{diag}(1,1)$. Note that\ $\chi ^{I}$\ with $I=1,2$
denote two massless scalar fields.

One obvious feature\ of this model is that the power $n$ can be a decimal.
In fact, it allows for explicit breaking of translational symmetry when $%
1/2\leq n<5/2$ and spontaneous breaking for $n>5/2$ \cite{Baggioli1904}. The
HR of this model has not been systematically studied before except for the
cases with $n=1$ \cite{Andrade1311} and $1/2$ \cite{Taylor1406}.

Let's build up the HJ formulation of this model. Using the metric%
\begin{equation}
ds^{2}=dr^{2}+\gamma _{ij}dx^{i}dx^{j},
\end{equation}%
one can obtain the Lagrangian from the action%
\begin{equation}
L=-\int_{\partial M}d^{3}x\sqrt{\gamma }\left[ R+6+\mathcal{K}^{2}-\mathcal{K%
}_{ij}\mathcal{K}^{ij}-V\left( G_{IJ}\dot{\chi}^{I}\dot{\chi}^{J}+\gamma
^{ij}\chi _{ij}\right) \right] .
\end{equation}%
It yields the canonical momenta%
\begin{eqnarray}
\pi ^{ij} &\equiv &\frac{\partial L}{\partial \dot{\gamma}_{ij}}=\sqrt{%
\gamma }(\mathcal{K}^{ij}-\mathcal{K}\gamma ^{ij}), \\
\pi _{I} &\equiv &\frac{\partial L}{\partial \dot{\chi}^{I}}=2\sqrt{\gamma }%
V^{\prime }\left( X\right) G_{IJ}\dot{\chi}^{J}.  \label{pichi}
\end{eqnarray}%
The Hamiltonian can be defined by a Legendre transformation of Lagrangian%
\begin{equation}
H\equiv \int_{\partial M}d^{3}x(\pi ^{ij}\dot{\gamma}_{ij}+\pi _{I}\dot{\chi}%
^{I})-L.  \label{HVn}
\end{equation}

We need to solve $\dot{\chi}^{I}$ from the nonlinear equation (\ref{pichi}).
This is difficult for a decimal $n$. Fortunately, in the holographic
application of this model, it is usually assumed implicitly%
\begin{equation}
V\left( G_{IJ}\dot{\chi}^{I}\dot{\chi}^{J}+\gamma ^{ij}\chi _{ij}\right)
=V\left( \chi \right) +V^{\prime }\left( \chi \right) G_{IJ}\dot{\chi}^{I}%
\dot{\chi}^{J}+\cdots ,  \label{Vexp}
\end{equation}%
where $\chi =\gamma ^{ij}\chi _{ij}$ and $\chi _{ij}=G_{IJ}\partial _{i}\chi
^{I}\partial _{j}\chi ^{J}$. Suppose that the asymptotic behavior of the
axions%
\begin{equation}
\chi ^{I}=\chi _{(0)}^{I}(x^{i})+e^{-\bar{\Delta}r}\chi
_{(1)}^{I}(x^{i})+\cdots .
\end{equation}%
The expansion (\ref{Vexp}) implies%
\begin{equation}
\bar{\Delta}>1.  \label{delta1}
\end{equation}%
Furthermore, one can find that the higher order terms in Eq. (\ref{Vexp}) do
not contribute to the counterterms, provided that%
\begin{equation}
2\left( n-1\right) +2\bar{\Delta}\geqslant 3.  \label{ndelta}
\end{equation}%
Here we will focus on the situation satisfied with Eq. (\ref{delta1}) and
Eq. (\ref{ndelta}). More complete analysis will be given in \cite{Wu2022}.

Now let's go back to Eq. (\ref{pichi}). With Eq. (\ref{Vexp}) in mind, it
can be solved as%
\begin{equation}
\dot{\chi}^{I}=\frac{G^{IJ}\pi _{J}}{2\sqrt{\gamma }V^{\prime }\left( \chi
\right) },
\end{equation}%
and the Hamiltonian has the form%
\begin{equation}
H=\int_{\partial M}d^{3}x\left[ \frac{1}{\sqrt{\gamma }}\left( \pi _{ij}\pi
^{ij}-\frac{1}{d-1}\pi ^{2}+\frac{1}{4}\bar{G}^{IJ}\pi _{I}\pi _{J}\right) +%
\sqrt{\gamma }L_{S}\right] ,  \label{HGbar}
\end{equation}%
where the effective coupling and the counteterm seed are%
\begin{equation}
\bar{G}^{IJ}=\frac{1}{V^{\prime }(\chi )}G^{IJ},\ L_{S}=R+6-V(\chi ).
\end{equation}

With the Hamiltonian in hand, we can derive the counterterm part of HJ
equation. The procedure is almost the same as in section A. The only thing
worth mentioning is that despite the appearance of non-trivial coupling $%
\bar{G}^{IJ}$ in the Hamiltonian, it disappears in its renormalized
counterpart, since here we have%
\begin{equation}
\frac{\delta S_{\mathrm{os}}}{\delta \Phi ^{I}}=\frac{\partial L}{\partial
\dot{\Phi}^{I}}=2\sqrt{\gamma }G_{IJ}V^{\prime }(\chi )\dot{\Phi}^{J}.
\end{equation}%
Furthermore, one can see that $\bar{G}^{IJ}$ and $V(\chi )$ are not
polynomials in general, which makes the ansatz generated below go beyond the
form (\ref{generalU}). Nevertheless, the variation identity (\ref{dbarU})
still holds while $\bar{k}$ may be a decimal. Collecting these facts, one
can find that the general algorithm is applicable to this model. In fact, it
is similar to previous two models since their coefficient functions $%
F(C_{(k_{i})})$ in the recursive equations are the same.

1. Generate ansatz%
\begin{eqnarray}
L_{A(2n)} &=&-V,\ L_{A(2)}=R, \\
L_{A(4n)} &=&4Y_{(2n)ij}Y_{(2n)}^{ij}+\frac{d+4n(n-1)}{1-d}U_{(2n)}^{2},
\end{eqnarray}%
\begin{eqnarray}
L_{A(2n+2)} &=&8Y_{(2n)ij}Y_{(2)}^{ij}\!\!+\!\!\ \bar{G}%
^{IJ}P_{I(2n)}P_{J(2n)}\!\!+\!\!\ \frac{2(d-2+2n)}{1-d}U_{(2n)}U_{(2)}, \\
L_{A(6n)} &=&8Y_{(2n)ij}Y_{(4n)}^{ij}+\frac{d+2n(4n-3)}{1-d}%
2U_{(2n)}U_{(4n)}.
\end{eqnarray}

2. Take variation%
\begin{equation}
Y_{(2n)ij}=-C_{(2n)}V^{\prime }\chi _{ij},\ P_{I(2n)}=2C_{(2n)}\nabla
_{i}\left( V^{\prime }\nabla ^{i}\chi _{I}\right) ,\ Y_{(2)ij}=C_{(2)}R_{ij},
\end{equation}%
\begin{equation}
Y_{(4n)ij}=C_{(4n)}C_{(2n)}^{2}\left\{ \frac{d+4n(n-1)}{1-d}2n\chi
^{2n-1}\chi _{ij}+8n^{2}\chi ^{2n-3}\left[ \left( n-1\right) \chi _{kl}\chi
^{kl}\chi _{ij}+\chi \chi _{i}^{k}\chi _{kj}\right] \right\} .
\end{equation}

3. Specify coefficients

We have $C_{(k_{i})}=-\frac{1}{2(d-k_{i})}$\ for $d\neq k_{i}$ or $-\frac{r}{%
2}$ for $d=k_{i}$. We have not written the concrete expression at each order
since $n$ has not been specified.

We collect $U_{(k_{i})}$ at all orders:%
\begin{eqnarray}
U_{(2n)} &=&C_{(2n)}L_{A(2n)},\ U_{(2)}=C_{(2)}L_{A(2)},  \label{Uaxion1} \\
U_{(4n)} &=&C_{(4n)}L_{A(4n)},\ U_{(2n+2)}\!\!=\!\!C_{(2n+2)}L_{A(2n+2)},\
U_{(6n)}=C_{(6n)}L_{A(6n)},  \label{Uaxion2}
\end{eqnarray}%
Since here $d=3$, the logarithmic divergence appears at $n=\frac{3}{2}$, $%
\frac{3}{4}$, and $\frac{1}{2}$. Note that we have sorted the orders based
on $\frac{1}{2}<n<1$. For $n=\frac{1}{2}$, $U_{(2)}$ and $U_{(4n)}$ appear
at the same order, and so do $U_{(2n+2)}$ and $U_{(6n)}$\footnote{%
In Ref. \cite{Taylor1406}, only $U_{(2n)}$ has been obtained. Using the
concrete black hole solution given in \cite{Baggioli2014}, we have double
checked that other counterterms are necessary for the finiteness of the
renormalized action \cite{Wu2022}.}. For $3/2\geq n\geq 1$, one can see that
only $U_{(2)}$ and $U_{(2n)}$ in Eq. (\ref{Uaxion1}) and Eq. (\ref{Uaxion2})
are relevant to real counterterms. For $n>3/2$, only $U_{(2)}$ remains to be
relevant.

\end{document}